\documentclass[lettersize,journal,review]{IEEEtran}

\usepackage{array}
    \newcolumntype{P}[1]{>{\raggedright\arraybackslash}p{#1}}
    \newcolumntype{M}[1]{>{\raggedright\arraybackslash}m{#1}}
    
\usepackage[caption=false,font=normalsize,labelfont=sf,textfont=sf]{subfig}
\usepackage{textcomp}
\usepackage{stfloats}
\usepackage{url}
\usepackage{verbatim}
\usepackage{graphicx}
\usepackage{cite}
\usepackage{enumitem}
\usepackage{multirow}
\usepackage{xtab}
    \xentrystretch{-0.2}
\usepackage{float}
\usepackage{xcolor}

\usepackage[numbers]{natbib}

\def\BibTeX{{\rm B\kern-.05em{\sc i\kern-.025em b}\kern-.08em
    T\kern-.1667em\lower.7ex\hbox{E}\kern-.125emX}}
\usepackage{balance}

\newcommand\citectx[1]{\citeauthor{#1} \cite{#1}}

\begin{document}
\title{Mask-Mediator-Wrapper architecture as a Data Mesh driver}

\author{
Juraj Dončević \IEEEmembership{Member, IEEE}, 
Krešimir Fertalj \IEEEmembership{Member, IEEE}, 
Mario Brčić \IEEEmembership{Member, IEEE}, 
Mihael Kovač \IEEEmembership{Member, IEEE}
\\ {\color{red}This work has been submitted to the IEEE for possible publication. Copyright may be transferred without notice, after which this version may no longer be accessible.}
\vspace{-18pt}
\thanks{J. Dončević, K. Fertalj, M. Brčić and M. Kovač are with University of Zagreb Faculty of Electrical Engineering and Computing, Zagreb, Croatia.}
}

\markboth{IEEE journal,~Vol.~X, No.~Y, September~2022 \\}{}

\maketitle

\begin{abstract}
The data mesh is a novel data management concept that emphasises the importance of a domain before technology. The concept is still in the early stages of development and many efforts to implement and use it are expected to have negative consequences for organizations due to a lack of technological guidelines and best practices. To mitigate the risk of negative outcomes this paper proposes the use of the mask-mediator-wrapper architecture as a data mesh driver. The mask-mediator-wrapper architecture provides a set of prefabricated configurable components that provide basic functionalities which a data mesh requires. This paper shows how the two concepts are compatible in terms of functionality, data modelling, evolvability and aligned capabilities. A mask-mediator-wrapper driven data mesh facilitates: low-risk adoption trials, rapid prototyping, standardization, and a guarantee of evolvability. 
\end{abstract}


\section{Introduction}
The landscape of data management is constantly changing. In just the recent two decades, there has been a great procession of technologies, formats, tools and systems that have shown promise in tackling data management. Data warehouses serving as analytical data stores were the common starting point. The situation became complicated with the introduction of NoSQL, as unstructured schemas, multitudes of data formats, and data federation became commonplace.

In an understandable effort to stay ahead of the competition, the industry more than adapted to these new ideas. The need for velocitous, voluminous and various data was stamped into the "3 Vs" slogan. With the demand now up to 8 Vs, a decade of data streaming, data lakes and data lakehouses is at a close, with many promising ideas ahead.

These are all exciting ideas to explore, and very few can really be considered obsolete. The problem with hitting the mark in data management is that the target is constantly moving, making data management an ongoing struggle susceptible to rapid paradigm shifts. The most recent of these is the introduction of the data mesh \cite{dehghani_data_2022}, leaving leading organizations \cite{nguonly_andrew_data_2021,lei_bo_data_2022,joshi_data_2021,databricks_data_2020} in a race to extract every ounce of benefit available to them from this new paradigm. The question is if all organizations can implement this shift quickly and stably enough with a positive outcome, as some might be forced to exhaustively restructure their entire data organization and provision. These organizations run the risk of implementing a half-baked solution they don't benefit from, ultimately quitting mid-shift with an unmanageable product. There is an implicit expectation that each organization will develop its own custom components to drive a data mesh, restricting inter-system composability, technologist migration among projects, and challenging evolvability. The impact of these problems is undeniable, with just 20 percent of analyses bringing value \cite{white_our_2019}, and the failure rate of data science projects at 87 percent \cite{venturebeat_what_2019}.

This papers shows that a data mesh can be driven by the MMW architecture, and contributes with a MMW-driven data mesh concept that enables:
\begin{itemize}
    \item \textit{low-risk adoption trials} for organizations that might not have data exactly \textit{at scale}
    \item \textit{rapid prototyping} to reduce the time required for implementation
    \item \textit{standardization} to reduce the cognitive load for technologists when switching projects and making the system composable
    \item \textit{a guarantee of evolvability} to alleviate technological changeability
\end{itemize}

Sections \ref{sec:data_mesh} and \ref{sec:mmw_arch} provide succinct overviews of the data mesh and MMW architecture concepts. Section \ref{sec:compatibility} discusses the compatibility of both concepts in terms of their functionalities, modelling abilities, architectural evolvability and data mesh capability coverage. This is demonstrated by constructing a hypothetical data mesh using the MMW architecture in Section~\ref{sec:compatibility}. Section \ref{sec:gains} discusses the benefits that can be gained by using the MMW architecture to drive a data mesh.

\section{The data mesh}\label{sec:data_mesh}

The data mesh is an enterprise data platform architecture that converges the ideas of \cite{dehghani_how_2019}:
\begin{itemize}
    \item \textit{distributed domain driven architecture}
    \item \textit{self-serve platform design}
    \item \textit{product thinking with data}
\end{itemize}

It is an alternative to the centralized data platform approach, where data is centrally managed and served through coupled extract-transform-load (ETL) processes. 

\begin{figure*}[]
    \centering
    \includegraphics[width=0.9\textwidth]{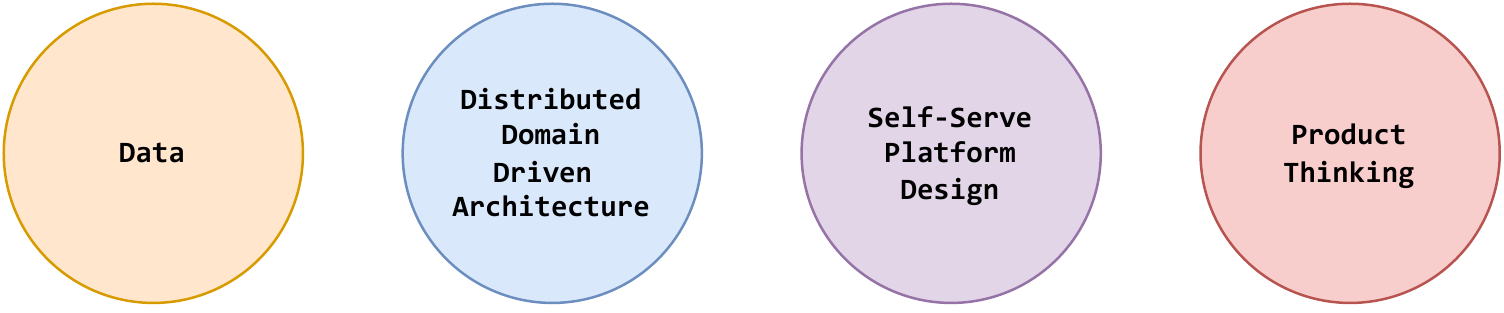}
    \caption{Convergence of ideas in the data mesh \cite{dehghani_how_2019}}
    \label{fig:paradigm_convergence}
\end{figure*}

The key organizational feature of the data mesh is that it arranges an organization's data into bounded contexts. This brings domain-driven design from operational systems into analytical systems. With this conceptual data federation in place, teams can more easily manage data in their domains, acquire domain-specific knowledge faster and handle data and new tasks with greater expertise. This means that responsibilities are distributed among teams along the bounded contexts rather than along mechanical functions (e.g. ingestion, cleansing, aggregation, serving) \cite{dehghani_how_2019}. Echoing the statement of \citectx{richards_fundamentals_2020} that in engineering architectures everything is a trade-off, the division by domains requires teams to be cross-functional. 

\begin{figure*}[]
    \centering
    \includegraphics[width=\textwidth]{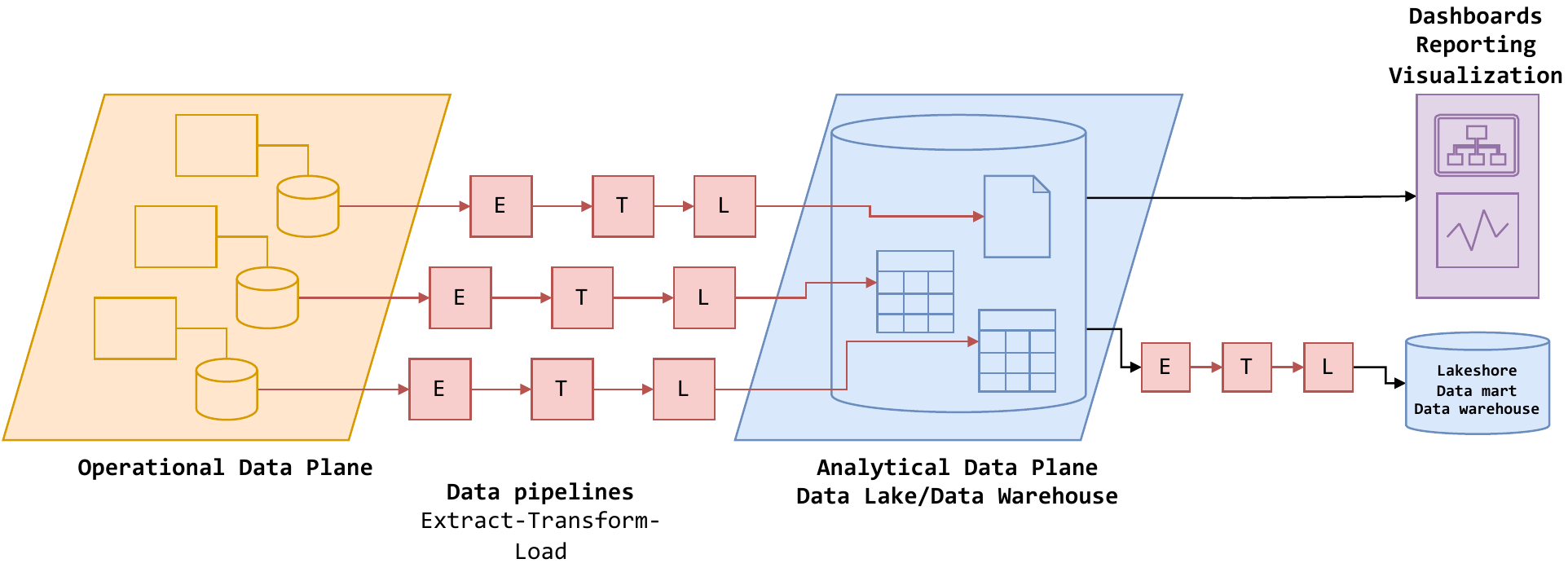}
    \caption{Centralized data management over planes (adapted from \cite{dehghani_how_2019})}
    \label{fig:centralized_architecture}
\end{figure*}

Each team creates a data pipeline to load domain data from the data infrastructure platform (DIP). Ideally, domains shouldn't overlap \cite{evans_domain-driven_2003}, but this is inevitable. In this case, a domain that uses data from another domain doesn't create its own separate data pipeline to the DIP, but correlates the data from an existing domain. Since data is considered a product, its quality is guaranteed by the other domain's team - the team owns the data. Verifying and maintaining data quality is simpler since each team manages their domain. 

The idea of the data mesh corresponds with the evolutionary architectures concept, as presented by \citectx{ford_building_2017}. Each domain can be considered an architectural quantum, which is according to the definition of \citectx{ford_building_2017} \textit{an independently deployable component with high functional cohesion, which includes all the structural elements required for the system to function properly}. Since cases where architectural quanta depend on each other are inevitable, this implies that the interfaces between domains must be designed technologically agnostic. Such interfacing is made easier since the data mesh domains have to follow global interoperability principles proscribed by federated computational governance. \citectx{dehghani_data_2022} even goes so far as specifying that data products (defined by domains) have input and output data ports as interface points. It can be noticed that the data mesh relies heavily on the ports-and-adapters pattern of the hexagonal architecture \cite{cockburn_hexagonal_2005}, which gives promise that the data mesh is a sound solution. Even Netflix, the proponents of the hexagonal architecture \cite{svrtan_ready_2020}, have been quick to admit that they find the data mesh compelling \cite{nguonly_andrew_data_2021, lei_bo_data_2022}.

\begin{figure}[H]
    \centering
    \includegraphics[width=0.45\textwidth]{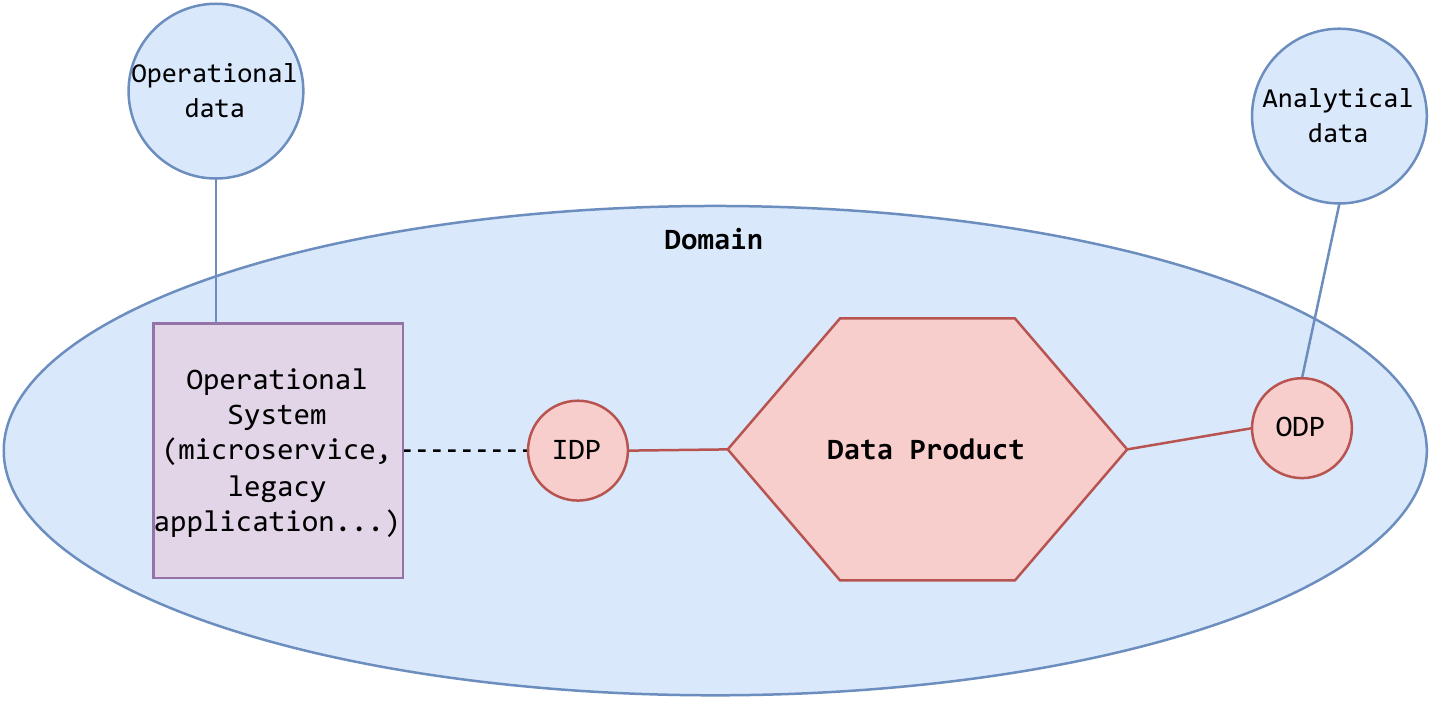}
    \caption{Data product's placement in the domain and interaction (IDP - input data port, ODP - output data port); adapted from \cite{dehghani_data_2020}}
    \label{fig:domain_hexagon}
\end{figure}

\begin{figure}[H]
    \centering
    \includegraphics[width=0.45\textwidth]{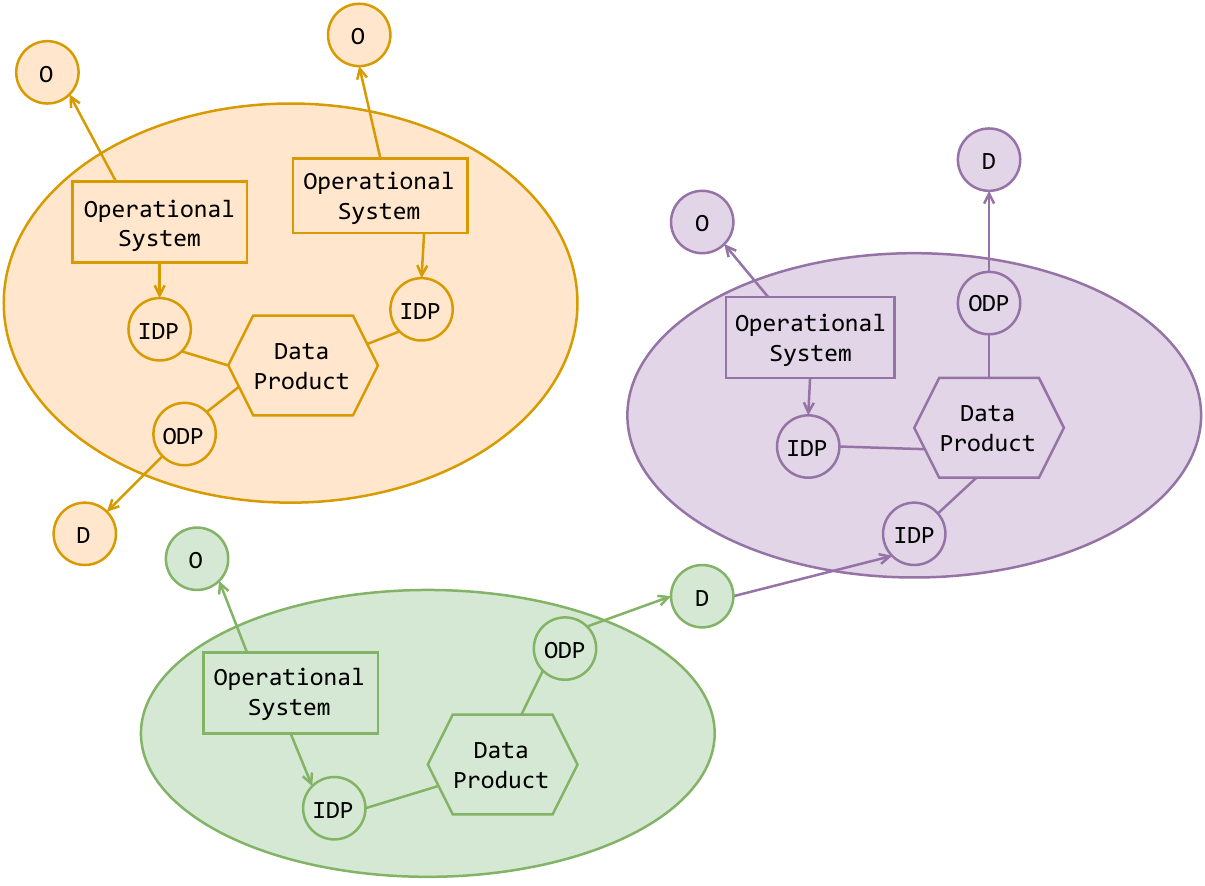}
    \caption{Multiple domains serving domain-oriented data (O - operational data; D - analytical data) \cite{dehghani_data_2020}}
    \label{fig:multiple_domains}
\end{figure}

In this domain-oriented architecture data is consumed by users (reporting, visualization, dashboards of Fig.~\ref{fig:centralized_architecture}), operational systems and analytical systems (other data products). The inter-connection possibilities are conceptually illustrated in Fig.~\ref{fig:multiple_domains}. It is important to note how some data products might depend on analytical data from other data products; by example of Fig.~\ref{fig:multiple_domains} the purple data product requires analytical data from the green domain. On the other hand, data products never depend on operational data from another domain; the purple data product didn't reach for the operational data from the green domain.

\begin{figure}[H]
    \centering
    \includegraphics[width=0.45\textwidth]{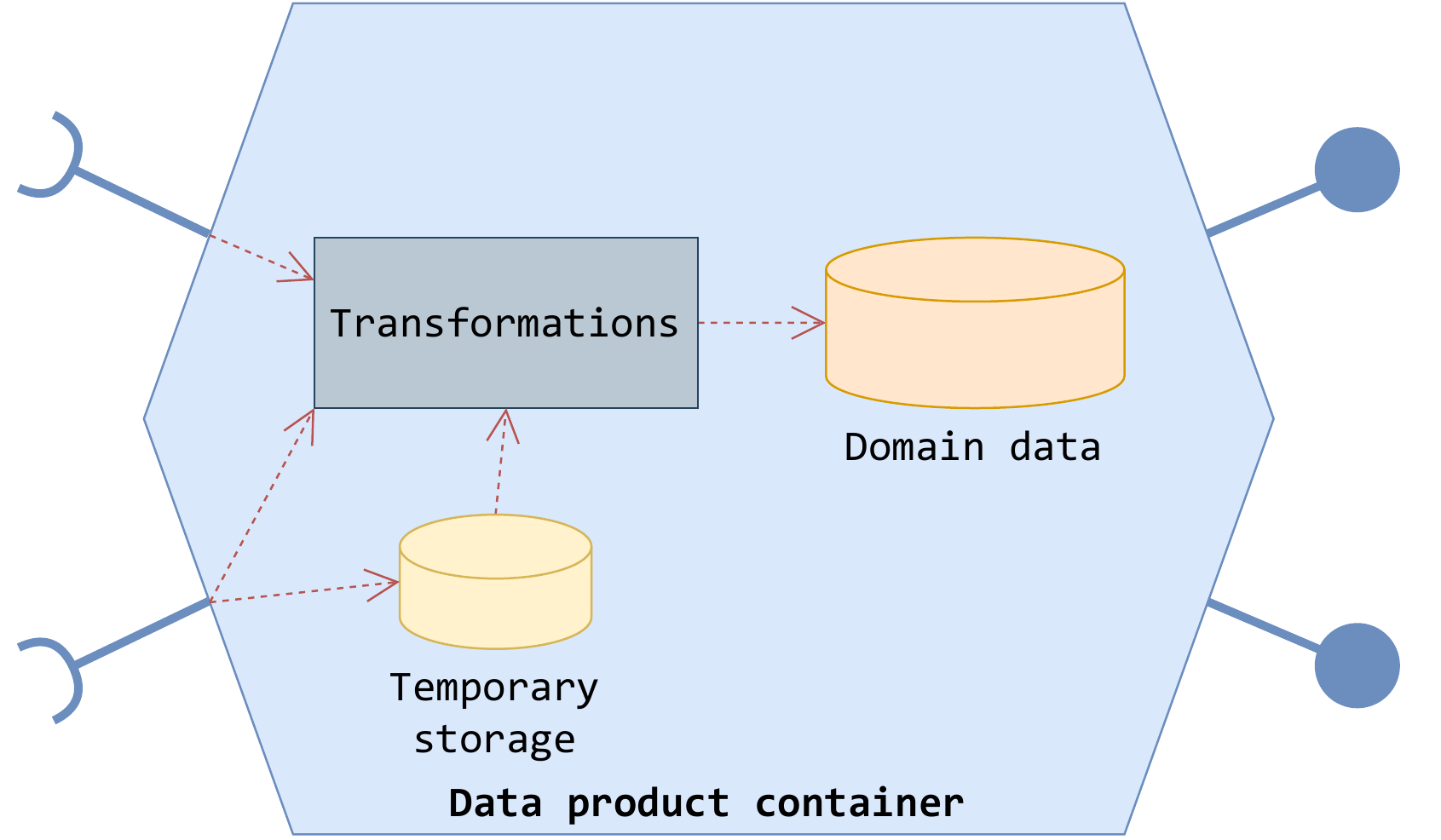}
    \caption{A data product container having a local domain data storage - data prepared for serving (adapted from \cite{dehghani_data_2022})}
    \label{fig:data_product_container}
\end{figure}

The DIP is seen as the primary data source location for the data mesh, containing data warehouses, data lakes or operational data stores as as data platforms \cite{waehner_heart_2022, desai_veeral_manage_2022}. On the other hand, \cite{dehghani_data_2022} mentions that there are other top-level data source archetypes in the data mesh such as collaborating operational systems or other data products. This raises a practical question of whether some data should remain physically close to the data product. In Fig.~\ref{fig:data_product_container} we present a distilled view on this by \citectx{dehghani_data_2022}, which proposes that a data product container can keep data locally. This local data is the domain data being served by the data product. The local data must be consumed from the input data ports and transformed to a valuable product. This all happens in the data product container.

The data mesh as a data platform enables \cite{dehghani_how_2019}:
\begin{itemize}
    \item Scalable polyglot big data storage
    \item Encryption for data at rest and in motion
    \item Data product versioning
    \item Data product schema
    \item Data product de-identification
    \item Unified data access control and logging
    \item Data pipeline implementation and orchestration
    \item Data product discovery, catalog registration and publishing
    \item Data governance and standardization
    \item Data product lineage
    \item Data product monitoring/alerting/log
    \item Data product quality metrics (collection and sharing)
    \item In memory data caching
    \item Federated identity management
    \item Compute and data locality
\end{itemize}

\section{The mask-mediator-wrapper architecture}\label{sec:mmw_arch}

\begin{figure}[H]
    \centering
    \includegraphics[width=0.35\textwidth]{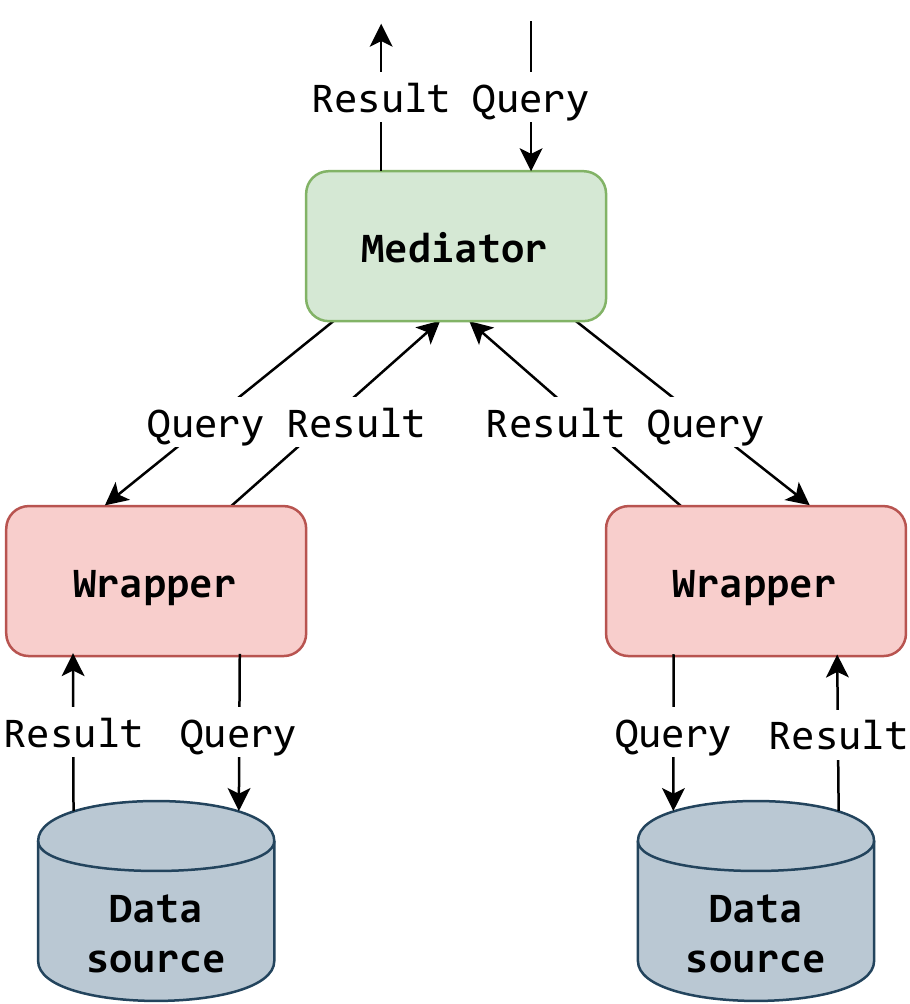}
    \caption{The mediator-wrapper architecture pattern}
    \label{fig:mw_architecture}
\end{figure}

The mask-mediator-wrapper (MMW) architecture, proposed by \citectx{doncevic_mask-mediator-wrapper_2022} is an extension of the already known mediator-wrapper (MW) architecture (Fig.~\ref{fig:mw_architecture}). The MW architecture is shown to underperform in terms of data representation, which is heavily leveraged in modern data management, so the MMW architecture proposes a new additional component type - \textit{the mask}. The work by \citectx{doncevic_mask-mediator-wrapper_2022} presents ground rules for MMW components, a quantitative flexibility analysis comparing the MMW with the MW architecture, and a case where a legacy store-preserving system could hypothetically be substituted by a MMW system. This hypothetical case study shows how the MMW architecture accommodates more than just data integration, but also other data management purposes.

The MMW architecture is comprised of three component types: \textit{mask}, \textit{mediator} and \textit{wrapper}. Each of these component types deals with a specific set of tasks and responsibilities: 
\begin{itemize}
    \item The wrapper is used to encapsulate a data source and provide a universal data source interface to the rest of the system. The wrapper allows the data source to be queried for both its data and metadata. The data sources can be a variety of storage solutions: relational and NoSQL databases, data lakes, or even web content.
    \item The mediator is used to transform and integrate data and metadata from wrappers or lower-tier mediators. Mediators can be composed to work in multiple tiers, each tier raising the level of abstraction or extracting a piece of integrated data and metadata.
    \item The mask is used to represent the data and metadata it acquires from a mediator in different formats. The mask allows a MMW system to be both materializing and virtualizing. A mask can provide a raw data acquisition interface, or a graphical user interface for reporting, or be used as a data warehouse loader.
\end{itemize}

\begin{figure}[H]
    \centering
    \includegraphics[width=0.45\textwidth]{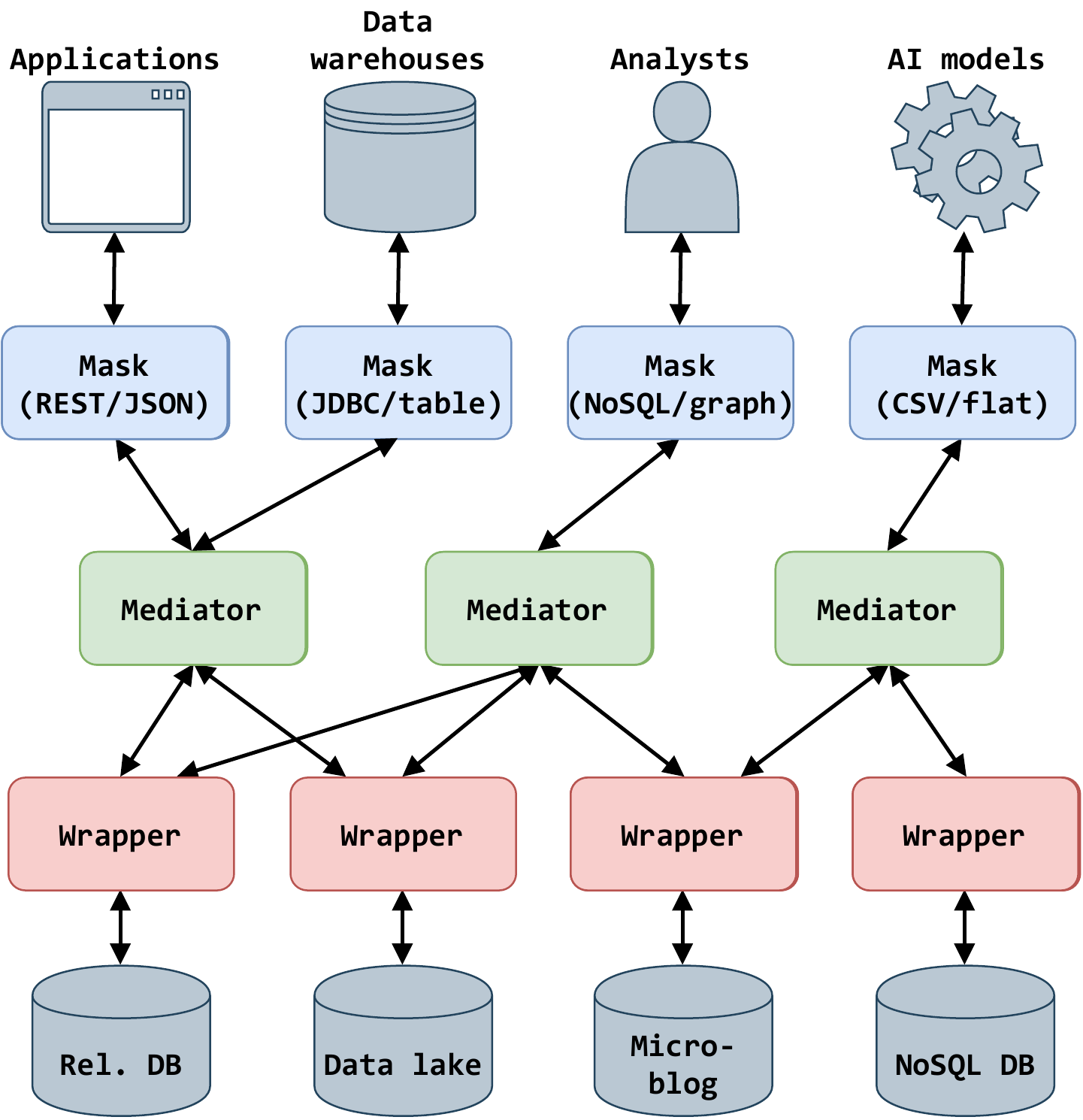}
    \caption{Mask-mediator-wrapper architecture example}
    \label{fig:mmw_architecture_example}
\end{figure}

The data and metadata that the components translate and exchange are schemas, queries and data. Schemas represent the organization of data, while queries over schemas are used to select which data is currently of interest. Despite this exchange, each component can be considered an architectural quantum as, by definition of \citectx{ford_building_2017}, each component is independently deployable and cohesive. High cohesion is a product of the refined separation of concerns among component types (set by the component type rules \cite{doncevic_mask-mediator-wrapper_2022}). Independent deployability is intrinsic to the componentization in the MMW architecture; for example a wrapper doesn't require other components to function properly, neither do mediators and masks (although they would present empty schemas and data sets).

Our currently ongoing research of translation in mediators is indicating that a loosely-relational data representation powered by bidirectional transformations presents a practical solution. \citectx{kleppmann_designing_2017} suggest that relational models generalize data very well, so they can be useful in a broad variety of cases beyond their originally intended scope of business data processing. The goal of the relational model was to hide this implementation detail behind a cleaner interface \cite{kleppmann_designing_2017}. This is fitting, since mediators should be able to present a schema merged from multiple heterogeneous sources. Bidirectionalization, as proposed by \citectx{asano_view-based_2018}, allows the mediators to have set rules that describe transformations for localization and globalization of schemas, queries and data. Additionaly, \citectx{asano_view-based_2018} even go so far as presenting an SQL-like language for declaring views (translated relations). These declarations can then be followed through certain rules to acquire bidirectional transformations of schemas, queries and data going into and out of the mediator; this generally comes down to query rewrites. In practice, this means that a generalized set of transformations can be created by an SQL-like declaration to manage schemas, queries and data in a mediator.

\section{Compatibility of concepts}\label{sec:compatibility}

Although at first glance the data mesh and the MMW architecture seem to be concepts from different time periods and paradigms, we consider them compatible. They cover similar problem areas, use similar concepts and provide a similar service. The MMW architecture is a much more primitive idea, than the data mesh. Whereas the data mesh focuses more on solving organizational problems (both human and data-related), the MMW tries to solve functional granularity problems and provide comprehensive reusable architectural quanta for data management. However, \citectx{dehghani_data_2022} and \citectx{doncevic_mask-mediator-wrapper_2022} stress that their concepts are domain and technology agnostic, giving their realizations the ability to prove compatible. The data mesh and the MMW architecture are orthogonal concepts, aiming to cover different problem areas in data management. Together with the promise of agnosticism, this provides the basis for the data mesh and the MMW architecture to be complementary ideas that can produce a greater effect when joined.

\subsection{Consume-Transform-Serve}
\begin{figure}[H]
    \centering
    \includegraphics[width=0.4\textwidth]{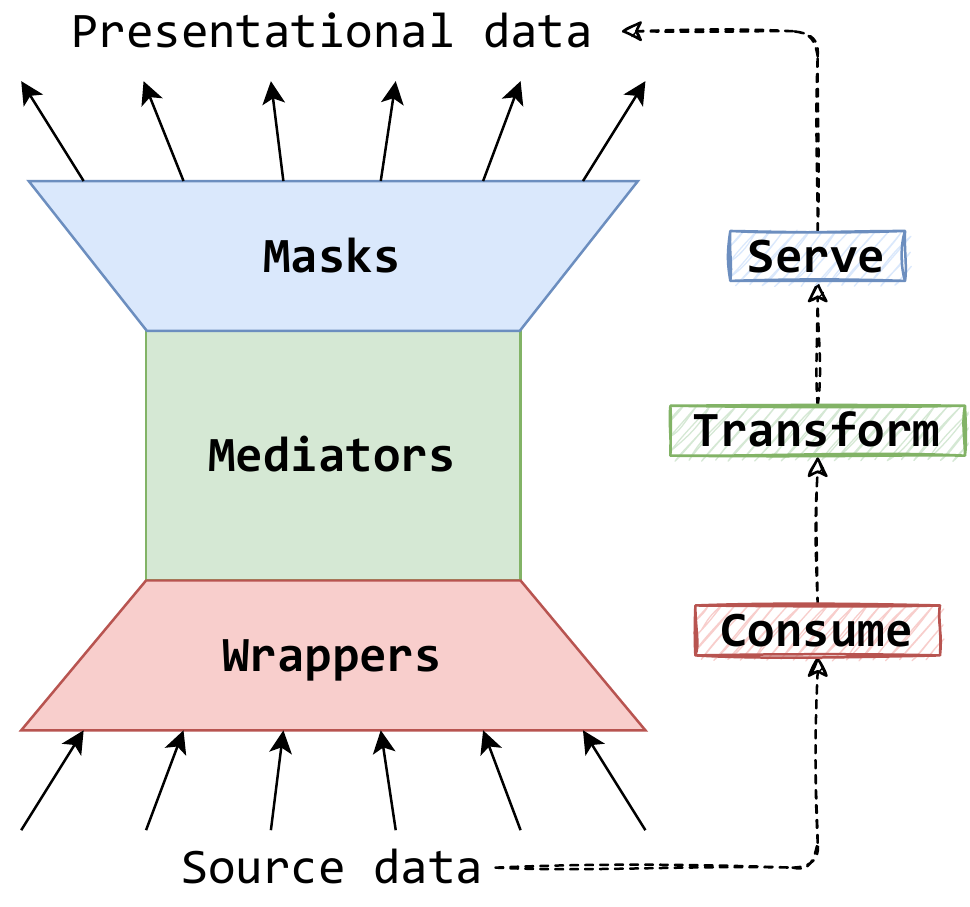}
    \caption{Consume-transform-serve concept relating to the MMW architecture (adapted from \citectx{doncevic_mask-mediator-wrapper_2022})}
    \label{fig:mmw_cts}
\end{figure}

A data mesh's data product is expected to autonomously consume, transform and serve data \citectx{dehghani_data_2022}. This is visible in Fig.~\ref{fig:data_product_container}, with the input ports used to consume data, transformations being done inside the container, and the data served through the output ports. In the MMW architecture (Fig.~\ref{fig:mmw_cts}), data consumption is driven by the wrappers, transformation is driven by mediators, and data serving is driven by masks. By providing consume-transform-serve capabilities, the components of a MMW architecture are capable of acting as a data product.

\subsection{Data models}
Data models in domain-driven design are instinctively thought of as constructs of classes or structs. This is because domain-driven design is usually observed in operational systems. In analytical systems this could be a pitfall, because the consumed data might be defined by different metamodels and it's expected that the data is served in polyglot form. \citectx{dehghani_data_2020} also recognizes this stating: \textit{Depending on the nature of the domain data and its consumption models, data can be served as events, batch files, relational tables, graphs, etc., while maintaining the same semantic}. Rather, domain-driven design should be considered at the most abstract level. \citectx{kleppmann_designing_2017} have explained why relational systems have stood the test of time: \textit{relational databases turned out to generalize very well, beyond their original scope of business data processing, to a broad variety of use cases}\cite{kleppmann_designing_2017}. The relational model, with its generalization abilities and time-enduring legacy in analytical systems, should be a good fit for the data mesh. Nothing prevents a domain from being described in a relational model - this is done on a regular basis in operational systems, especially when developing micro-services.

A relational model could also be used in an MMW architecture. The use of a tabular relational-like model in mediators has already been demonstrated by \citectx{asano_view-based_2018}. The relational model's generalization abilities \cite{kleppmann_designing_2017} is a confident sign that this should be the data model of choice. Support for polyglot access would also be significantly easier to research and implement, as there are already numerous implementations of relational mapping to other models. This is the focal point of the MMW architectures capability to be domain-agnostic, while enabling modeling, processing and sharing across the organization, just as stipulated by \citectx{dehghani_data_2022}.

\subsection{Evolutionary compatibility} \label{sec:evolutionary_compatibility}
\citectx{dehghani_data_2022} declares that: \textit{"it’s only appropriate to [...] leave the specific implementation details and technology to be refined and built over time. [...] any specific implementation design or tooling suggestion will be simply outdated by the time you get to read this book"}. Technology is considered capricious in this research area, and it is now common sense to avoid any technological dependency when discussing architectures and concepts in software engineering. Technological changeability is one of the driving values of evolutionary architectures \cite{ford_building_2017}, from which \citectx{dehghani_data_2022} derives the suggestions that are made about the data mesh's hypothetical implementation. Essentially, a data mesh should be evolvable, and the MMW architecture should follow suit if it is to be compatible.

The MMW architecture fits the common dimensions of evolvability as follows:
\begin{itemize}
    \item \textbf{Technical}\\
    The inner components of the mask, mediator and wrapper component types are finely grained \cite{doncevic_mask-mediator-wrapper_2022} and their core functionalities can be separated by interfacing from technologically changeable inner components. This allows the MMW components to be adapted to different technologies (e.g. in terms of communication protocols and data formats). This decoupling also allows components supporting newer or previously unsupported technologies to be developed quickly.
    \item \textbf{Data} \\
    Data shared across a MMW system is both metadata and data, described in a generalized manner to begin with. Changes to the core metamodels are expected to be infrequent, but to minimize the effects of changes, the definitions can be placed in a shared library.
    \item \textbf{Security} \\
    Security related to authentication, authorization, confidentiality and data integrity for communication requirements can be interfaced along-side communication protocols and data formats. Security in terms of logical operations (e.g. sharing confidential system information or sending unsafe data) can be assured through a common library for each component type as well as defining a standard set of inter-component exchanges.
    \item \textbf{Operational/System}\\
    Since each MMW system component is an architectural quantum, mapping the components to an existing infrastructure is very flexible.
\end{itemize}

The evolvability of the MMW architecture allows it to uphold the data mesh's set of conventions that promote interoperability between different technologies, which guarantee its longevity alongside the data mesh concept.

\subsection{Capability coverage} 
A set of capabilities the data mesh offers was mentioned in Section~\ref{sec:data_mesh}. The MMW can cover these fifteen capabilities as follows in Table~\ref{tab:capability_coverage}.

\tablecaption{Data mesh capability coverage by the MMW architecture}\label{tab:capability_coverage}

\begin{xtabular}[]{|P{0.12\textwidth}|P{0.3\textwidth}|}

\hline
\textbf{Scalable polyglot big data storage}                          & The MMW architecture allows usage of heterogeneous data source and heterogeneous representation. \\ \hline
\textbf{Encryption for data at rest and in motion}                   & MMW components can be interface their communication with encryption protocols. Encryption of static data comes down to the encryption of local data storage. \\ \hline
\textbf{Data product versioning}                                     & Data products can be versioned over a metamodel. As a hypothetical example on a relational model, tables are grouped in schemas, and schemas represent specific versions of data products. \\ \hline
\textbf{Data product schema}                                         & A data product schema is provided by default in each component, because queries are declared over schemata. \\ \hline
\textbf{Data product de-identification}                              & De-identification can be done during data transformation in mediators, or via specialized instructions for wrappers. \\ \hline
\textbf{Unified data access control and logging}                     & Data access can be controlled in all MMW component types. Masks can also provide access control via their applications. \\ \hline
\textbf{Data pipeline implementation and orchestration}              & Data pipelines are implemented by composing the mask, mediator and wrapper components to consume, transform and serve data. \\ \hline
\textbf{Data product discovery, catalog registration and publishing} & Data product discovery is enabled by examining schemata provided by components. This process is simplified, since all component types have common interfaces. \\ \hline
\textbf{Data governance and standardization}                         & Data is governance is federalized because MMW components can work in separate groups. Standardization is provided by the MMW compomnents' standard interfaces. \\ \hline
\textbf{Data product lineage}                                        & Lineage can be overseen by looking into mediator transformations, and mask and wrapper translations. \\ \hline
\textbf{Data product monitoring/alerting/log}                        & Each MMW component can be deployed along with a monitoring application. Logging is expected to be component-level in all components types. Alerting interfaces can also be put in place as part of the core component or the monitoring application. \\ \hline
\textbf{Data product quality metrics (collection and sharing)}       & Quality metrics can be shared as a part of the schema metadata. This allows quality metrics to be defined for each data product version separately. \\ \hline
\textbf{In-memory data caching}                                      & In-memory data caching can be implemented for each component type to optimize query response times. \\ \hline
\textbf{Federated identity management}                               & Identity management is a part of the infrastructure capabilities, but the MMW architecture doesn't prohibit or discourage such cases. \\ \hline
\textbf{Compute and data locality}                                   & Data can be locally transformed in mediators and placed in local storage using materializing masks; then to be consumed by a wrapper on request and served (see Section~\ref{sec:construction}). \\ \hline
\end{xtabular}
\medskip

\subsubsection{Constructing a data mesh platform}\label{sec:construction}
\begin{figure}[H]
    \centering
    \includegraphics[width=0.45\textwidth]{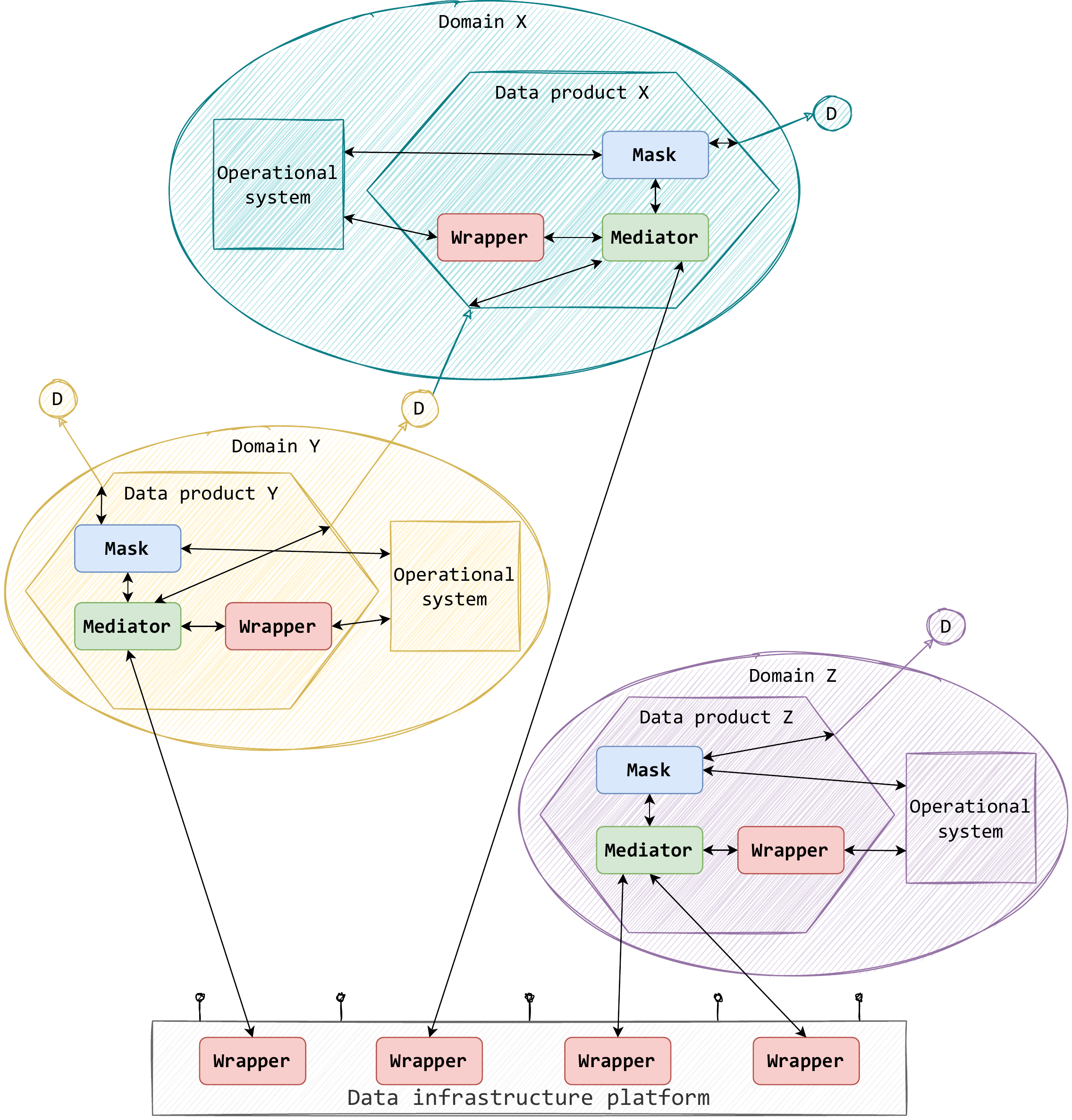}
    \caption{MMW architecture used to drive a data mesh}
    \label{fig:mmw_data_mesh}
\end{figure}

\hfill\\
A generic example of a MMW driven data mesh is shown in Fig.~\ref{fig:mmw_data_mesh} for some nondescript domains \textit{X}, \textit{Y} and \textit{Z}. The example is similar the one given in Fig.~\ref{fig:multiple_domains} to give the reader some bearing. Each domain being covered by an operational system and a data product serving analytical data. Data products consume data from their domain's operational system or the DIP.

A DIP should solve the need for duplicating data pipelines, storage and streaming infrastructure \cite{dehghani_data_2022}, making it the platform for uniformed data access. This platform can be fittingly driven by multiple wrappers. Each wrapper can cover a single data source, or there might be a case where multiple wrappers cover a single data source (e.g. load balancing, or wrapper specialization at runtime). This provides standardized access to data sources and the DIP itself.

Data products access data from the DIP using their mediators, since the data and metadata are already consumed and translated by wrappers. A data product may access data consumed by multiple wrappers (like in the example of \textit{Domain Z} in Fig.~\ref{fig:mmw_data_mesh}). A data product accesses data from their domain's operational system using a wrapper. The operational data is consumed through a wrapper either connecting to an operational database or the operational system's API. The mediator combines and transforms the consumed data into a data product. The data product is served through a mask. Masks can also serve data to their domain's operational system. To reduce intermediate architectural steps, data products can acquire data from other domains by connecting those data products' mediators directly. This is exemplified for Domains \textit{X} and \textit{Y} in Fig.~\ref{fig:mmw_data_mesh}, where the \textit{Domain X} mediator in effect requests data from the \textit{Domain Y} mediator. Direct access to these mediators from quanta that are not mediators in other data products should be prohibited by policies of the federated computational governance.

\subsubsection{Constructing a data product with localized storage}
\begin{figure}[H]
    \centering
    \includegraphics[width=0.45\textwidth]{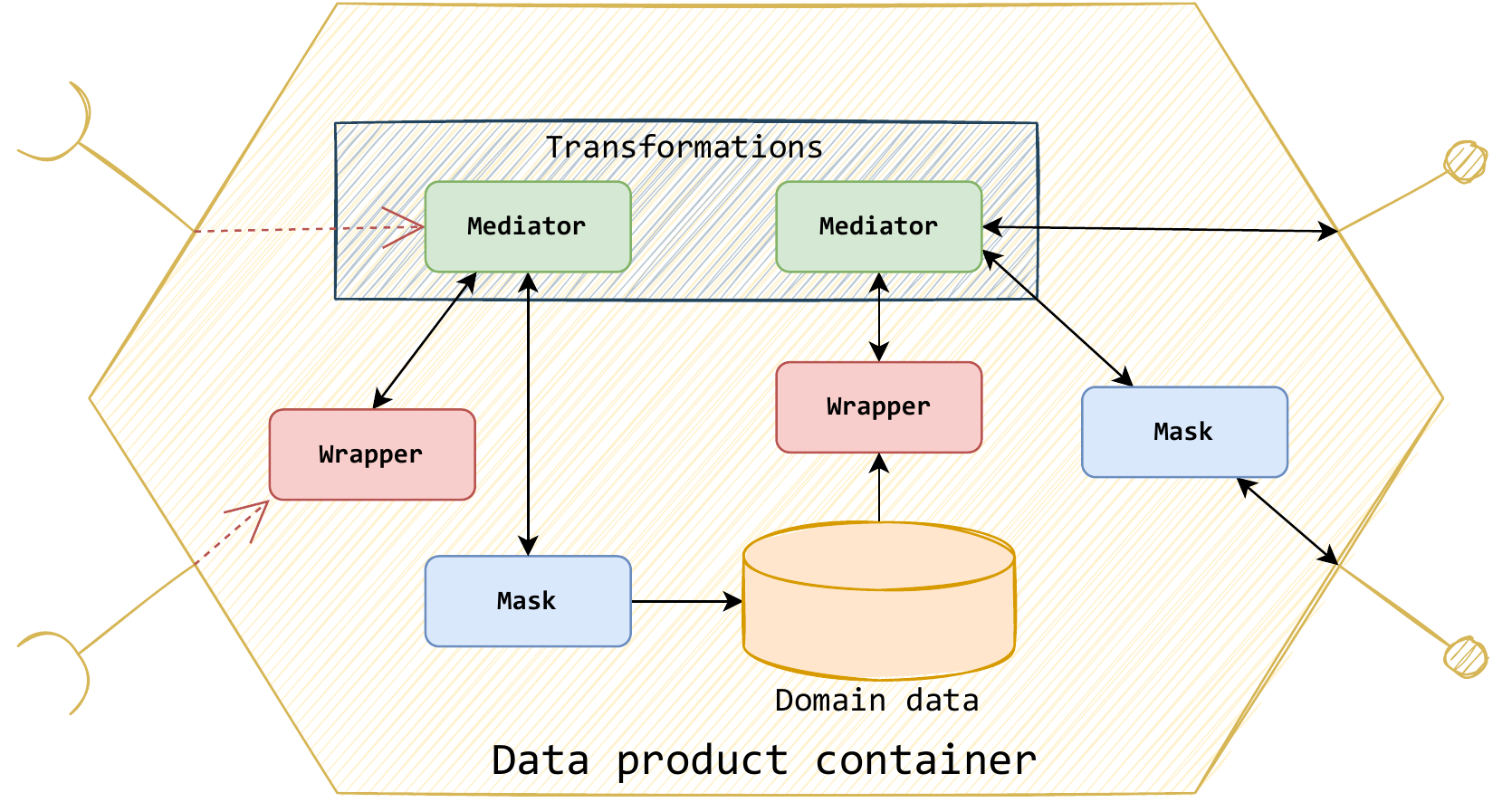}
    \caption{MMW architecture used to emulate a data product with localized domain data}
    \label{fig:mmw_data_product}
\end{figure}

\hfill\\
Analogously to the product container presentation in Fig.~\ref{fig:data_product_container}, Fig.~\ref{fig:mmw_data_product} displays how it could be driven by MMW components if a data product contained a local domain data storage. Data is acquired either as a data product from another domain directly by a mediator or by consumption of the domain operational data via a wrapper. The (left-hand) mediator transforms the data in preparation for it to be stored as a product in the domain data storage. A mask then materializes the data in the domain data storage. This persisted data can then be set into motion by a wrapper's consumption. The data is originally requested by a (right-hand) mediator. Theoretically a mask can connect directly to a wrapper \cite{doncevic_mask-mediator-wrapper_2022}, but not omitting the mediator allows some manoeuvre space for additional transformations before serving the data by a mask. The data that is served to data products from other domains by the mediator responsible for the finalizing transformations (in this case the right-hand side one).

The aforementioned cases presented in Fig.~\ref{fig:mmw_data_mesh} and Fig.~\ref{fig:mmw_data_product} demonstrate how the MMW architecture can drive the entire data mesh boilerplate. The cases also demonstrate that custom-built adapters are not required to fit the data mesh's functionality.

\section{What are the gains?}\label{sec:gains}

\subsection{Low-risk adoption trials}
The data mesh promises getting value out of data \textit{at a scale}. Determining that scale currently remains a rule of thumb, and it's unlikely that a metric will be proposed until a larger number of organizations try to adopt the data mesh - failing or succeeding. The point remains that the data mesh is intended for organizations that store various and voluminous data. Would a small local convenience store or a local accountant's office benefit from a data mesh? Probably not. Would a streaming service or an online market service benefit from a data mesh? Very likely.

Those organizations that are considered between the two exemplified groups are at most risk. Adopting a new analytical system, in a novel architecture, and built from scratch is not a simple operation. Many human and economic factors can lead the adoption astray. The system might be too complex to use in a simpler business environment, it might increase latency without providing any tangible analytical flexibility, and it might not even be developed to product-level, leaving the analytical capabilities of the organization in disarray. Even worse, if engineers unwittingly use the elephant migration anti-pattern, they might find themselves with an unfinished data mesh and a partially-dismantled legacy system. This is all without mentioning the time, human and financial resources invested. These cases are expected to be common, as just 20 percent of analytic insights are expected to deliver business outcomes through 2022 \cite{white_our_2019}, and 87 percent of data science projects never make it into production \cite{venturebeat_what_2019}.

Organizations can greatly reduce the risk adopting a data mesh by setting up a trial run by using the MMW architecture. The risk is reduced by using the MMW architecture in the following ways:
\begin{itemize}
    \item \textit{Development failure}\\
    MMW components require development only if specialized components are required. This is an edge case if wrappers and masks are not developed for certain data source types or data representation. Components are largely expected to just be acquired, deployed and configured. There is no extensive coding required.
    \item \textit{Loss of large resource investment} \\
    Since there is no extensive coding required when using the MMW architecture, a small technical or development team could prepare a \textbf{demonstrative data mesh in short time}. The financial investment can boil down to the price of additional infrastructure (if needed), and that could be constrained to a cloud service so no additional hardware acquisition is required.
    \item \textit{Deteriorated business usability} \\
    The MMW architecture doesn't require the legacy analytical system to be dismantled, so a possible deterioration of service is limited to the time frame of the adoption trial run. If the data mesh is found unsuitable for the organization, then it can be easily dismantled and the analytical system reintroduced. 
\end{itemize}

\subsection{Rapid prototyping}
The MMW architecture allows the data mesh to be rapidly prototyped when initiating analytical capabilities in an organization. The entire setup of the system is realistic. The initial step is to deploy and configure MMW components to drive a prototypical data mesh. When requirements are fully distilled over the prototype, the MMW-driven data products can then be substituted piecemeal with permanently developed ones. The DIP is migrated incrementally as permanent data products are developed and deployed.

Rapid prototyping via the MMW architecture allows the data mesh to be expeditiously deployed in an organization's environment, so benefits can be reaped as soon as possible. It enables early problem detection, business process and organizational structure alignment, as well as bringing an increase in business product value sooner.

\subsection{Evolvability}
It is expected that components driving the data mesh form an evolvable architecture. In reality, the software development industry is known to omit beneficial architectural and design system properties for the sake of reaching a minimum-viable-product and creating profit quickly. While the consideration of evolvability is seen as obvious, it is questionable how many software development projects will continue to follow evolvability principles throughout their life-cycles. Another problem is that senior developers, designers or architects tend to stack technologies instead of components and quanta when working on systems.

It was shown in Section~\ref{sec:evolutionary_compatibility} that the MMW architecture is evolvable. Evidently, a composition of evolvable architectural quanta transiently makes their composition evolvable, hence MMW components can be used as a set of building blocks for a data mesh \textbf{to guarantee evolvability}. 

\subsection{Standardization}
Since the data mesh concept proposes no explicit implementation or use of technology, it's expected that organizations will implement data meshes as custom-built platforms. According to Conway's law \cite{conway_how_1968}, organizations will implement the data mesh according to their specific knowledge-base; using specific design patterns and technologies used by their technologists. Because of this, it is expected that each data mesh implementation will be significantly different from another. The discrepancy will lead to a lack of standardization, which can cause: a lack in composability of multiple data meshes in case of mergers, increased learning curve for newer technologists, a significant learning curve for experienced technologists switching projects (this diminishes their existing skill-set and making it largely unusable), or an implementation of a data mesh far removed from the original concept with detrimental effects on the organization. 

\citectx{dehghani_data_2022} stresses the importance of lowering the cognitive load of developers by using experiences, languages and APIs that are easy to learn as a starting point. With the MMW architecture a step further can be taken, and a standardization of architectural quanta created with which data meshes are regularly developed. This further lowers the cognitive load, allowing technologists to be calmly migrated among projects. Standardization benefits composability of systems, the composing systems use the same interfaces to function. It can be stated that standardization lowers degrees of implementation freedom, but it brings the beneficial effect of keeping projects close to the original data mesh concept.

\section{Conclusion}

Despite the data mesh being a promising concept, it is still in its infancy and, as is with any such new concept, lacks concerted best practices and standards. These will undoubtedly arrive as real-world development experience is gathered by early adopters. Potential data mesh adopters may come from a multitude of business areas and come in different organizational sizes, so it would be beneficial for them to run a trial data mesh before committing to full adoption. Standardization of a data mesh allows the system to be composable with other data mesh implementations, and allows technologist to be migrated from project to project with a minimal cognitive load. The use of a MMW architecture to drive a data mesh addresses all of those concerns.

The MMW architecture and data mesh are orthogonal concepts - the data mesh concerns itself with the organization of data management, while the MMW architecture solves functional granularity problems for quanta in data management. The concepts are compatible in terms of consume-transfer-serve functionalities, data modelling, evolvability and coverage of capabilities. The compatibility and orthogonality of the data mesh and MMW architecture concepts means they can be used side-by-side. The demonstrative examples have shown that a MMW driven data mesh is feasible.

The benefits of a MMW driven data mesh are: the ability run low risk adoption trials, the ability to develop a data mesh using rapid prototyping, guaranteed evolvability, and standardization. Both early and late adopters can find these benefits useful. Organizations at the edge of managing large scale data can run trials to see if a data mesh suits them. Organizations that require a data mesh to be developed quickly can use the MMW architecture to rapidly prototype it. Organizations concerned with standardization and evolvability can adopt the MMW driven data mesh as a complete solution.

Looking to future research, the MMW architecture might also be used to drive other data management architectures (e.g. data hub, data fabric, data spoke). If these suspicions are shown true, as in the case of the data mesh, further research might also explore the ability for the MMW architecture to drive migrations to other architectures. This includes state-of-the-art architectures. The MMW architecture might prove to be a long standing driver that enables technologists and their organizations to execute generational migrations between data management architectures. This could provide standardized and low-risk adoption of state-of-the-art architectures, while allowing the technological environment to evolve.

\bibliographystyle{IEEEtranN}
\bibliography{IEEEabrv,main}

\end{document}